\documentclass[twocolumn]{article}
\usepackage{graphicx}
\usepackage{authblk}
\usepackage[left=2cm,right=2cm]{geometry}
\usepackage{caption}
\usepackage{subcaption}
\usepackage[title]{appendix}
\usepackage{multirow}
\def\code#1{\texttt{#1}}

\title{Validation of NSFsim as a Grad-Shafranov Equilibrium Solver at DIII-D}
\author[1]{Randall Clark} 
\author[2]{Maxim Nurgaliev} 
\author[2]{Eduard Khayrutdinov} 
\author[2]{Georgy Subbotin}
\author[3]{Anders Welander}
\author[1]{Dmitri M. Orlov}
\affil[1]{Center for Energy Research, University of California San Diego, La Jolla, CA, 92093}
\affil[2]{Next Step Fusion S.a.r.l., Luxembourg}
\affil[3]{General Atomics, San Diego, CA, 92186}

\date{}

\begin{document}


\twocolumn[
  \begin{@twocolumnfalse}
    \maketitle
    \begin{abstract}
      Plasma shape is a significant factor that must be considered for any Fusion Pilot Plant (FPP) as it has significant consequences for plasma stability and core confinement. A new simulator, NSFsim, has been developed based on a historically successful code, DINA \cite{khayrutdinov1993studies}, offering tools to simulate both transport and plasma shape. Specifically, NSFsim is a free boundary equilibrium and transport solver and has been configured to match the properties of the DIII-D tokamak. This paper is focused on validating the Grad-Shafranov (GS) solver of NSFsim by analyzing its ability to recreate the plasma shape, the poloidal flux distribution, and the measurements of the simulated diagnostic signals originating from flux loops and magnetic probes in DIII-D. Five different plasma shapes are simulated to show the robustness of NSFsim to different plasma conditions; these shapes are Lower Single Null (LSN), Upper Single Null (USN), Double Null (DN), Inner Wall Limited (IWL), and Negative Triangularity (NT). The NSFsim results are compared against real measured signals, magnetic profile fits from EFIT \cite{lao1985reconstruction}, and another plasma equilibrium simulator, GSevolve \cite{welander2019closed}. EFIT reconstructions of shots are readily available at DIII-D, but GSevolve was manually ran by us to provide simulation data to compare against.
    \end{abstract}
  \end{@twocolumnfalse}
]

\section{Introduction}
Plasma shape has a long history of significant importance to improving fusion capability in tokamaks. DIII-D's elongated D shape offers hotter and denser stable plasma compared to earlier shapes like circular or elongated hourglass shapes historically used. Careful attention has been placed on the triangularity of the plasma, the location of X-points, and how their resulting strike points. \cite{lazarus1994role}

Negative triangularity plasmas have gained a lot of interest in the fusion community. One study showed that by using negative triangularity plasma they could both avoid ELMs that occur in the ELMy H-mode regimes while maintaining a level of plasma performance comparable to an H-mode \cite{nelson2023robust}. Another study has gone so far as to design an entire FPP based on negative triangularity \cite{rutherford2024manta}.

It is clear that the need for robust shape controls will be needed for any future Fusion Power Plant (FPP) no matter what shape it will end up being. It is also possible that a machine learning-based controller will be needed to control the plasma in this environment where the level of diagnostic measurement is significantly reduced compared to those in current research tokamaks. Reinforcement learning-based controllers have already been shown be capable of controlling plasma shape \cite{degrave2022magnetic}.

The simulation tool discussed in this paper, NSFsim, has been developed to analyze plasma shapes and offer tools to physicists to design new controllers to control plasma shape. The platform NSFsim has been developed on is a machine learning-friendly environment, offering easy access to Python-based machine-learning tools. In this paper, NSFsim has been shown to be a valid magnetics simulator of the DIII-D tokamak, enabling future work for tools to be developed on NSFsim to be compatible with the DIII-D tokamak.
\section{The Simulation Methods}
\subsection{NSFsim}
\subsubsection{An Introduction to NSFsim}
\begin{figure*}[!ht]
\centering
\includegraphics[width=0.8\textwidth]{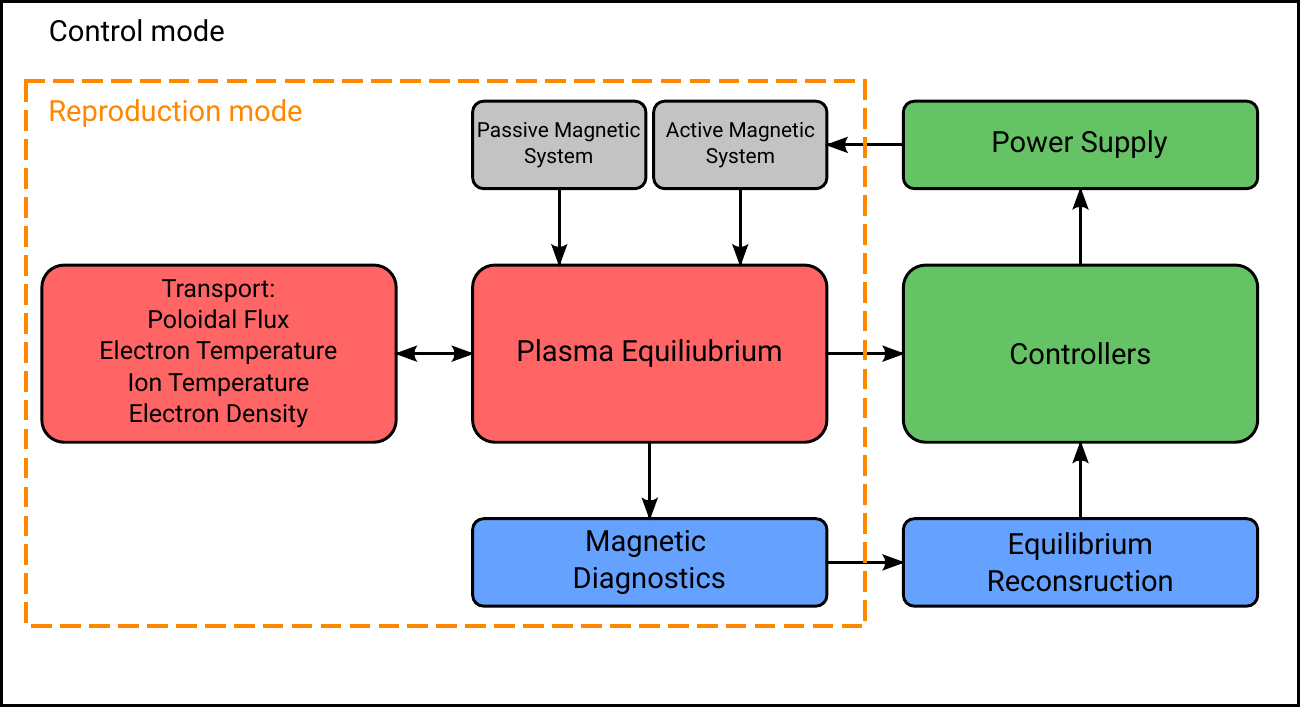}
\caption{This is a simplified overview of the NSFsim plant model of operation. The plasma flux surfaces are evolved in time in tandem with the transport equations. Simulated diagnostic signals are generated and these are used to control the externally applied magnetic fields, particularly the E- and F-coils at DIII-D. While there are different tools in place related to different control systems, for the purposes of a one-to-one validation with EFIT data, we loaded archived coil currents from DIII-D shots for feed-forward control, also referred to as reproduction mode in NSFsim, in place of an active feedback controller that is typically needed for operation.}
\label{DINAPlantModel}
\end{figure*}

NSFsim is a time-depended free boundary Grad-Shafranov (GS) solver coupled with transport equations that is based on DINA code \cite{khayrutdinov1993studies} (see section \ref{NumericSection} for more details). It has been developed as a platform for plasma shape analysis and control tasks. NSFsim was built in a framework that makes it a very accessible tool to combine with machine learning techniques as well. A general overview of the plant model can be seen in Figure(\ref{DINAPlantModel}). 

To calculate the magnetic equilibrium, the GS equation is needed:
\begin{equation}\label{GS_Equation}
\Delta^{*} \psi = -\mu_{0}R^{2}p'-ff',
\end{equation}
where $\psi$ is poloidal flux, $R$ is the major radius, $p$ is the pressure, $f(\psi)= RB_{\varphi}$, and $'$ denotes a derivative with respect to $\psi$. Additionally, the toroidal current density equation is also needed to for the magnetic equilibrium calculation:
\begin{equation}\label{Current_Density_NSF}
j_{\varphi} = Rp'+ff'/(\mu_{0}R)
\end{equation}

The transport equations that NSFsim uses, specially the diffusion equation for poloidal magnetic flux in 1D, the electron energy balance equation in 1D, the ion energy balance equation in 1D, and the particle transport equation in 1D, are described in detail in \cite{khayrutdinov1993studies}. 

The implementation of NSFsim in DIII-D (or any other tokamak for that matter) required the following items:
\begin{enumerate}
    \item[1)] The geometry of the tokamak (shape, position, size)
    \item[2)] The electrical properties of the magnetic coils (number of turns, resistivity)
    \item[3)] A vacuum vessel model – divided into filaments with finite thickness and resistivity
    \item[4)] A magnetic diagnostics setup describing the location and properties of all the sensors
    \item[5)] The limiter contour
    \item[6)] A defined boundary and initial condition before beginning any magnetics task
\end{enumerate}

For a description of the numerical solution to the plant model and its system of equations, see section \ref{NumericSection}

\subsubsection{Numerical Methods of NSFsim} \label{NumericSection}
NSFsim inherits many numerical methods originally developed for DINA code \cite{khayrutdinov1993studies}. DINA code is a powerful tool that has been used on many tokamaks for different applications \cite{khayrutdinov2001comparing,favez2001comparing,tamai2002runaway,lukash2007influence,kim2009full,miyamoto2014inter,xue2015simulation,xue2019hot}. For the convenience of the reader, we provide a brief explanation of the numerical scheme involved in running NSFsim.

Firstly, the GS equation is solved using two methods:
\begin{enumerate}
    \item[1)] Buneman’s methods, with a five-point stencil scheme, is used to
determine plasma boundary coordinates $\{r,z\}_b$ on rectangular grid. \cite{buneman1969compact}
    \item[2)] Inverse variable technique is used to determine metric coefficients
required for transport equations on polar grid ${\rho,\theta}$. \cite{degtyarev1985inverse}
\end{enumerate}

Then, to solve the transport equations, a homogeneous conservative difference approximation
scheme is used. This scheme is obtained using the integro-interpolation method \cite{Samarskiy1975} and is implemented using the flow variant of the sweep method for difference problems with highly varying coefficients \cite{degtyarev1969flow}. Energy balance equations for electrons and ions are solved using a matrix version of this method developed specifically for highly coupled equations.

The code has two global cycles of iterative processes that implement convergence in
non-linearities. The first cycle includes modules for calculating plasma equilibrium, the diffusion
of poloidal magnetic flux, and currents in external conducting structures. After the convergence of the first iteration loop, the transition to the second iterative loop occurs. It corrects the term, which is used again in the first iteration loop to clarify the current density distribution of the plasma.

The code comes in the form of unique universal solvers of direct and inverse problems. Its
uniqueness is in the ability to vary the number and quality of accounting parameters depending
on the task set before the calculations without the need to recompile the product. Functions for
adjusting the primary state in various ways (both by setting the coefficients and by specifying
profiles for the derivatives of magnetic pressure and flux) are used. Equations for several
variations of maintaining plasma stability have been introduced into the code. The equations of
the circuits and the equations of the field at a point have been clarified to improve the accuracy
of restoring electromagnetic diagnostic data. The possibility of parametric accounting of component heating
and adjustment for accounting for the isotope mixture of the working gas ions has been
introduced as well. 

NSFsim integrates advanced features such as energy and electron core transport through scalings or external first-principle codes, impurity transport, neutral particle simulation, ion and electron heating, and accelerated electron dynamics. Additionally, dedicated modules for simulating disruption events and breakdown processes enhance its fidelity across diverse plasma scenarios.

\subsubsection{NSFsim as an Environment for Reinforcement Learning}
NSFsim retains the core capabilities of the DINA code \cite{khayrutdinov1993studies}, while also providing enhanced interfaces tailored for machine learning applications. Its computational model is fine-tuned for increased computational speed and efficiency by customizing the simulation processes to match specific machine learning input/output requirements while adhering to preset constraints. This design streamlines NSFsim for rapid model training and experimentation with neural networks.

The code has undergone significant refinement, including optimization of plasma geometric parameter calculations to enhance efficiency. A balance was achieved between the speed and accuracy of plasma initialization at the first step. Adjustments were also made to the recalculation algorithm for synthetic electromagnetic diagnostic matrices, aimed at improving precision.

To increase flexibility in simulating the power supply characteristics of different devices, refinements were made to circuit equation calculations, supporting a variety of operational modes. These include active coils controlled by voltages with current recalculations in passive structures and options to switch coils between active and passive modes during simulations. This setup allows NSFsim to accurately simulate DIII-D’s patch panel, specifying which coils are active and which are passive.

NSFsim is written in \code{FORTRAN90}. To enable the use of our NSFsim for deep reinforcement learning (RL) applications, we have integrated it with \code{Python} via the \code{ctypes} library. This integration supports Gymnasium API \cite{towers2024gymnasium}, a widely used standard for RL training applications. The \code{FORTRAN} shared library exposes call functions to run a required type of simulation.    through \code{Python} interface passes necessary inputs and collects outputs. The environment class is specifically designed to match the structure and requirements of the Gymnasium API, which standardizes the interface for environments used in reinforcement learning.  The mapping of physical functions to standard API allows users to easily integrate the environment with popular deep RL libraries and frameworks, such as Stable-Baselines3 \cite{Raffin2021SB3} and RLlib \cite{liang2018rllib}.

The shared library works in one thread. To help RL algorithms easily collect “experience”, also known as episodes (actions and states), our \code{Python} environment can be instantiated in multiple processes, up to the number of available CPUs on a server. Each process runs an independent instance of the simulator, allowing multiple agents to interact with and learn from the environment simultaneously. This approach accelerates the collection of experiences for the agent, enhancing the efficiency of experience replay mechanisms used in deep RL.

The integration of our \code{FORTRAN}-based simulator with Python and the Gymnasium API broadens the accessibility of our simulation environment for reinforcement learning applications. This setup, which leverages the distributed nature of CPU-bound processes, offers a practical solution for efficient training of RL models, making it ideal for use in controlling real devices in a simulated-to-real pipeline.

\subsubsection{The Low-$\beta$ Limit}
It is important to note that simulations in this work are focused on low-$\beta$ shots. While NSFsim is both a GS and transport solver, our aim in this paper is to focus solely on the magnetics tasks (things like shape control and magnetic diagnostic simulation). For this reason, we chose to simulate shots that had smaller $\beta$ values to decouple the transport behavior from the magnetic behavior and minimize the influence of transport uncertainties on magnetic equilibrium. In the limit that $\beta$ becomes large, the transport effects on magnetic tasks become non-negligible; this arises because the $dp/d\Psi$ term in the toroidal current equation, also known as the bootstrap current, grows with $\beta$, and any mismatch in pressure will result in growing errors in the magnetic equilibrium. Therefore, to validate the magnetic capability of NSFsim, shots with low-$\beta$ were selected for validation in the Data Validation section. In the future validation of transport simulations and the ability of NSFsim to work as a plant simulator will be demonstrated.

\subsection{GSevolve}
\begin{figure}[!ht]
\includegraphics[width=.48\textwidth]{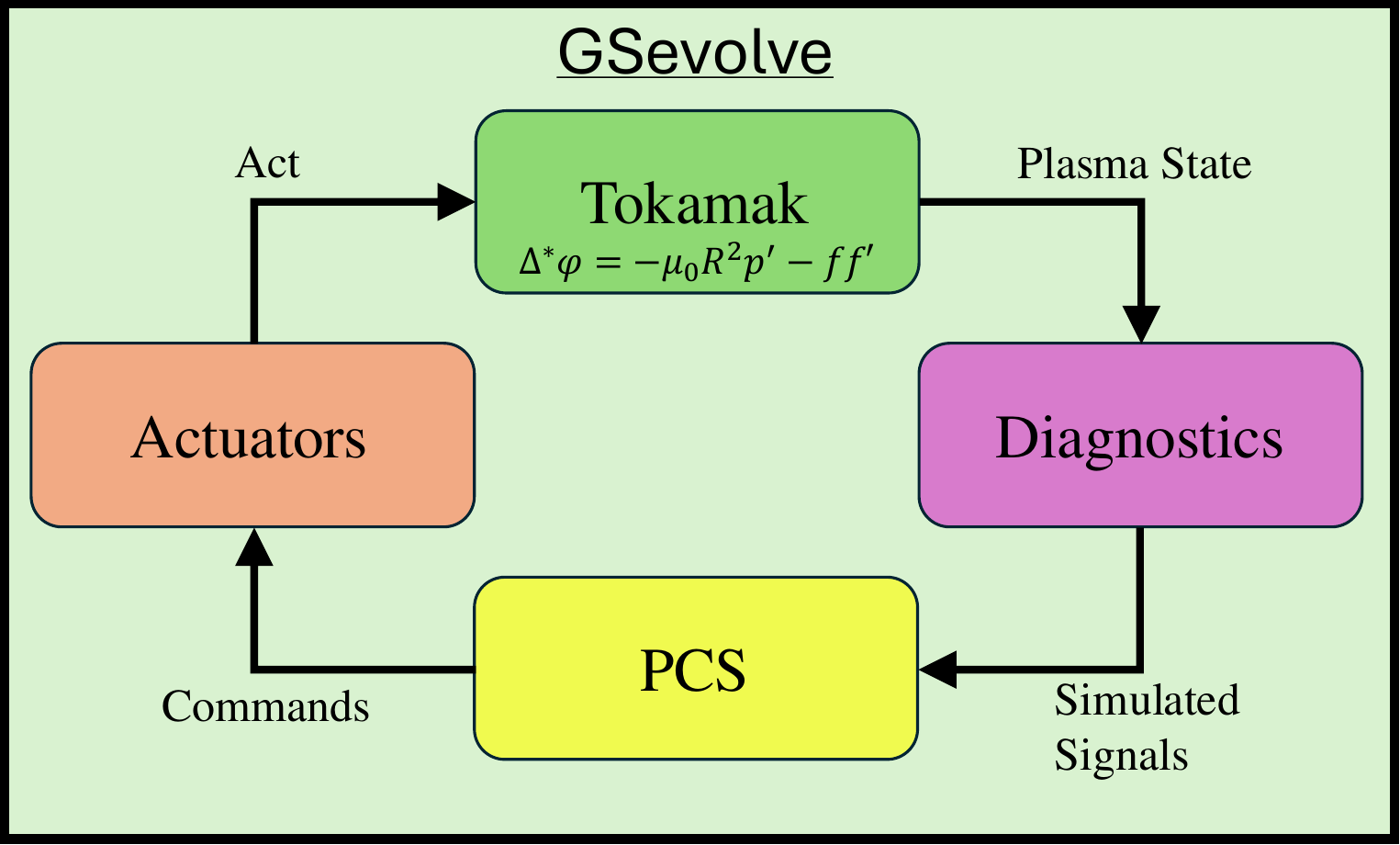}
\caption{This is the feedback loop for how GSevolve simulates the plant model of DIII-D. The tokamak begins with some initial conditions of the plasma which is evolved forward by GSevolve (in the green block called “Tokamak”). GSevolve will perform one iteration and forward the plasma state to the diagnostics where diagnostic sensors are simulated and noise is added. These simulated signals are sent into the Plasma Control System (PCS), which then outputs commands to a model of the DIII-D power supplies (the Actuators). The actuators send their voltage outputs to GSevolve to perform the next iteration to the plasma. This plant model enables researchers to test and experiment with new control algorithms in the PCS. \cite{welander2019closed}}
\label{GSevolvePlantModel}
\end{figure}

GSevolve is a nonlinear free-boundary simulation code that solves the GS equation (\ref{GS_Equation}), evolving the equilibrium over time, and can account for pressure and current profiles; we have used GSevolve to create simulations of the same shots that were simulated in NSFsim, doing so gives us an additional set of data to compare against. While NSFsim uses the measured coil currents and EFIT data to simulate and compare with old shots, GSevolve uses the archived PCS targets and shot configuration to control the plasma as it once was during the live shot.

GSevolve is able to simulate the evolution of current continuously from open to closed field lines which enables it to simulate the full discharge of a plasma from breakdown through plasma termination. It has been used as a tool for plasma control design and has been implemented in a plant model (see Figure \ref{GSevolvePlantModel}) replicating the plasma control workflow at DIII-D during a live shot. GSevolve is a PCSSP-compliant software with a successful application at both DIII-D and EAST \cite{welander2019closed}.

GSevolve's plasma simulation exists in the Tokamak block of the DIII-D Simulink model and works by iterating through the equilibrium evolution. It begins by calculating the applied flux from the given vessel and coil currents using known mutual inductances; then the plasma flux is calculated by first calculating the toroidal current density using equation(\ref{Current_Density_NSF}).

By breaking up the poloidal cross-section into a grid of rectangular cells that carry current equal to the surface integral of the toroidal current density equation (\ref{Current_Density_NSF}), the poloidal magnetic flux can be calculated using the known mutual inductances. The calculation of current density is done by modeling the profiles of $p'$ and $ff'$ as splined polynomial functions of normalized poloidal flux whose polynomial coefficients can vary freely. With these terms GSevolve can evolve the plasma equilibrium while taking into account the plasma response, the change in the plasma current when flux changes. See reference \cite{welander2019closed} for a more in-depth review of how GSevolve works.
 
In our GSevolve simulations, the existing PCS along with the built-in Isoflux algorithm were used to perform shape control and to take the plasma through ramp up, flat top, and ramp down all while controlling vertical instability. Isoflux works by creating a basis shape from user-chosen shape parameters and defines control points from this basis shape; one or more F-coils will be assigned to a control point to keep the poloidal flux the same at all control points (maintaining the desired plasma boundary). In the simulations that are shown in the next section, five experimental shots were recreated in GSevolve by taking the archived hardware setup and PCS settings (including the target shape parameters) from the real shot and running GSevolve with these conditions. This setup will generate a plasma very similar to the one experimentally measured and provide us with a data set to validate NSFsim's simulation of the same shot. \cite{ferron1998real,eldon2020high}

\subsection{Differences Between the Simulators and their Setup}
GSevovle is a well-established code at DIII-D built to connect with the plasma control system for controller development and prediction of plasma response; therefore, its data is useful for validating the accuracy of NSFsim data. However, a comparison between GSevolve and NSFsim is not a direct comparison because GSevolve uses a built-in controller to model a live shot in the DIII-D PCS system with feedback control while NSFsim runs in the reproduction mode, utilizing recorded coil currents for feed-forward control and uses an artificial vertical control scheme that adds in a small amount of poloidal flux (less than $\pm2.5$\% of the total poloidal flux) needed to prevent vertical instability. It is for these reasons that GSevolve will serve as a useful baseline for a quality simulator to compare against rather than a benchmark to declare one simulator more or less accurate than another.
\section{Data Validation}
\begin{figure*}[!ht]
\centering
\includegraphics[width=16cm,height=12cm,keepaspectratio]{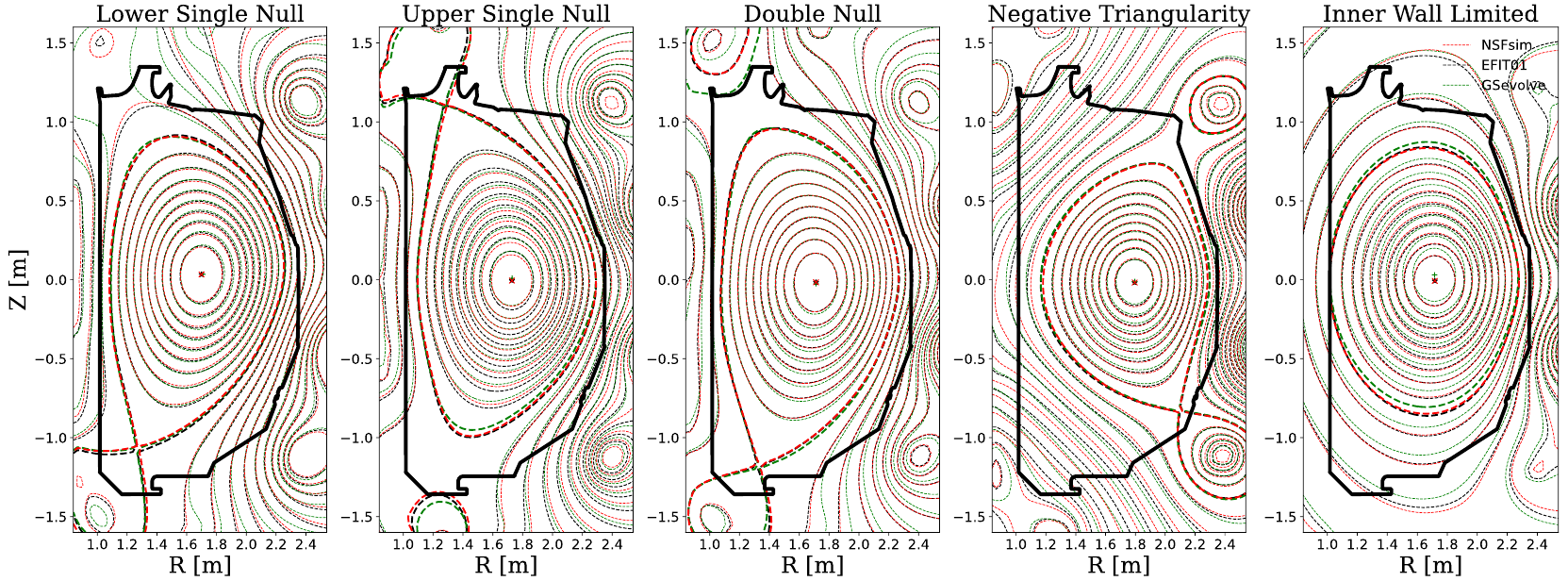}
\caption{Contour maps of the poloidal flux for each of the five plasma shapes, LSN 182450, USN 182626, DN 182444, NT 194667, and IWL 182392. The plots show strong agreement between the three sets of data regarding the plasma boundary and poloidal flux contours. The time slices for each contour plot were taken at different times, see Table \ref{Contour_Table} for the recorded time of each plot.}
\label{Contours}
\end{figure*}

\subsection{Poloidal Flux Contour Plots}
The primary goal of NSFsim as it is currently developed is to solve the GS equation, finding the force balance that defines the shapes and contours of the plasma. For this reason, the contours of GSevolve, EFIT01, and NSFsim for each of the five shapes were plotted out at select times.

\begin{table}[!ht]
\begin{tabular}{||c c c c||} 
 \hline
 Shape & Time Slice & NSFsim Start & NSFsim End \\ [0.5ex] 
 \hline\hline
 LSN & 3000 ms & 1100 ms & 4100 ms\\ 
 \hline
 USN & 4000 ms & 1100 ms & 4100 ms\\
 \hline
 DN & 3000 ms & 2500 ms & 5000 ms\\
 \hline
 NT & 2000 ms & 1100 ms & 3761 ms\\
 \hline
 IWL & 1500 ms & 1100 ms & 4002 ms\\ [1ex] 
 \hline
\end{tabular}
\caption{This table lists out the time slice that the contours in Figure \ref{Contours} were taken at. It also lists out the time during the shot NSFsim began at and the time it concluded simulation relative to the shot.}
\label{Contour_Table}
\end{table}

The times were chosen to be varied and to make sure the contour plot was as readable as possible. Figure \ref{Contours} provides a somewhat qualitative estimate of "goodness" in the simulation, meaning that the shape is a proper match and the contours follow their expected lines. For a more thorough analysis of the quality of the simulation as a function of time, the time series data of poloidal flux at the magnetic axis and plasma boundary are examined in the next section.

\subsection{Time Series Poloidal Flux Data}
\begin{figure}[!ht]
\includegraphics[width=.48\textwidth]{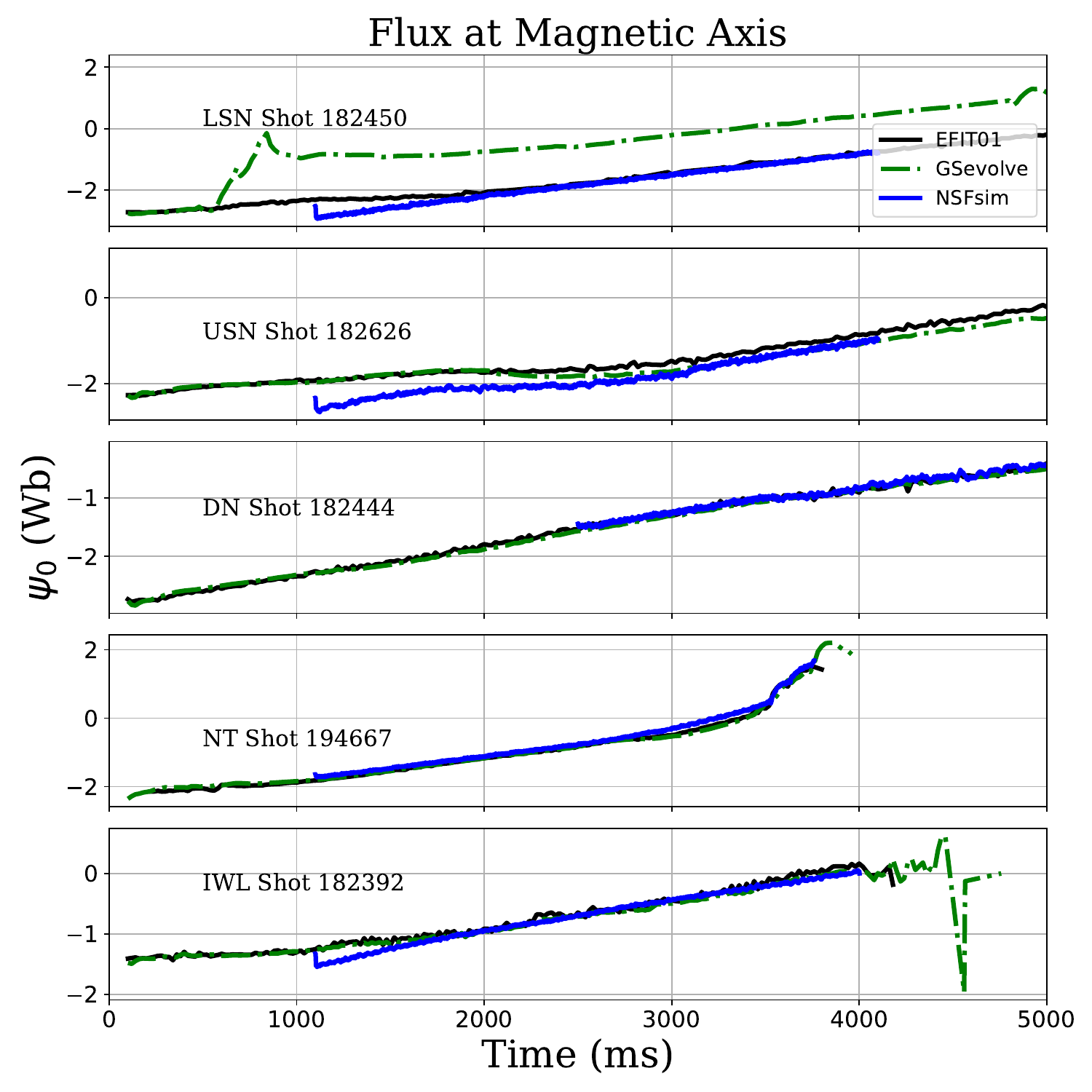}
\caption{These graphs depict the value of the poloidal flux at the magnetic axis for both the NSFsim and GSevolve simulations.}
\label{Flux_zero}
\end{figure}

\begin{figure}[!ht]
\includegraphics[width=.48\textwidth]{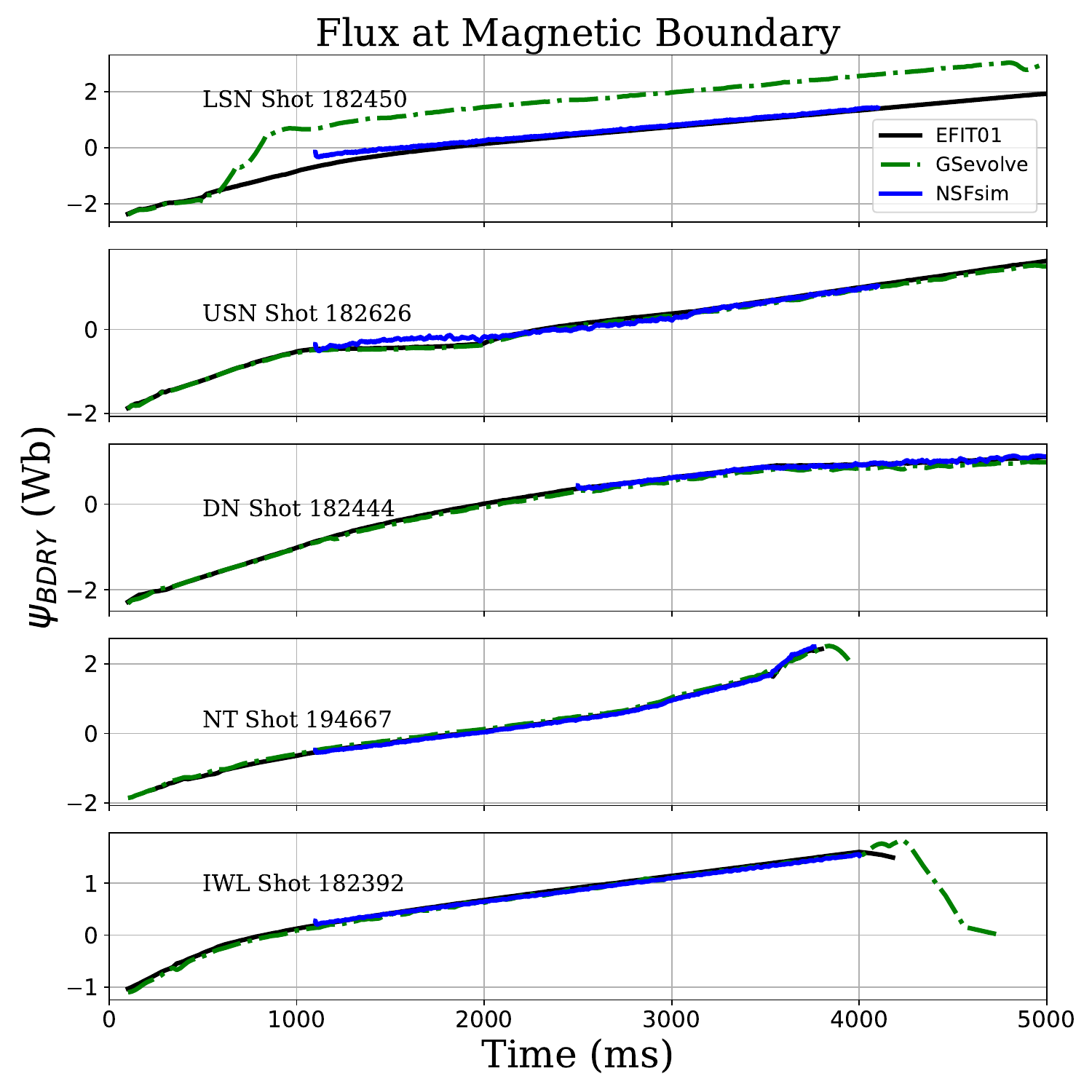}
\caption{These graphs depict the value of the poloidal flux at the plasma boundary for both the NSFsim and GSevolve simulations.}
\label{Flux_BDRY}
\end{figure}

Two points of significant interest are the magnetic axis and plasma boundary, particularly for NSFsim's ability to predict plasma shape. Figures \ref{Flux_zero} and \ref{Flux_BDRY} display the ability for NSFsim to accurately predict the poloidal flux at DIII-D throughout the flat top of a shot in these two regions. As shown in Table \ref{Flux_RMSE_table}, the RMSE of the poloidal flux is quite low and is comparable to GSevolve. It's important to note that GSevolve hits an unavoidable unstable spot early on in shot 182450 and accrues an offset to the poloidal flux that impacts its absolute value of poloidal flux; as shape is controlled in the PCS based off of relative fluxes, the error offset doesn't negatively impact control in the GSevovle environment.

NSFsim starts from the middle of the shot, at the beginning of the flat top, to avoid the heavy influence of transport effects during ramp-up; the formation and expansion of plasma create a volatile environment that makes our simulation prone to large errors in the estimation of the plasma pressure and induced toroidal currents in passive structures. Starting the simulation mid-shot leads to an abrupt change in poloidal flux values during the first few steps, as shown in Figures \ref{Flux_zero} and \ref{Flux_BDRY}. During the next few hundred milliseconds, the loss of poloidal flux is restored and the solution converges to experimental values. The most significant difference is observed in USN shot 182626 before 3 s due to the presence of high injected power for achieving QH-mode \cite{burrell2002quiescent}. To model such regimes, a more detailed consideration of transport and current generation processes is required. The implementation of codes and models for this is a high priority for the further development of NSFsim.


\begin{table}[!ht]
\begin{tabular}{||c c c ||} 
 \hline
 Boundary RMSE & NSFsim & GSevolve \\ [0.5ex] 
 \hline\hline
 LSN & 0.15 Wb & 1.1 Wb\\ 
 \hline
 USN & 0.11 Wb & 0.061 Wb \\
 \hline
 DN & 0.031 Wb & 0.078 Wb \\
 \hline
 NT & 0.037 Wb & 0.059 Wb \\
 \hline
 IWL & 0.039 Wb & 0.054 Wb \\ [1ex] 
 \hline \hline 
 Magnetic Axis RMSE & NSFsim & GSevolve \\ [0.5ex] 
 \hline\hline
 LSN & 0.25 Wb & 1.3 Wb \\ 
 \hline
 USN & 0.36 Wb & 0.15 Wb \\
 \hline
 DN & 0.049 Wb & 0.054 Wb \\
 \hline
 NT & 0.11 Wb & 0.079 Wb \\
 \hline
 IWL & 0.12 Wb & 0.063 Wb \\ [1.0ex] 
    \hline
\end{tabular}
\caption{These two tables list out the root mean square error for the poloidal flux at the plasma boundary and the magnetic axis as simulated by NSFsim and GSevolve.}
\label{Flux_RMSE_table}
\end{table}

\subsection{Magnetic Sensors}
\begin{figure}[!ht]
\includegraphics[width=.48\textwidth]{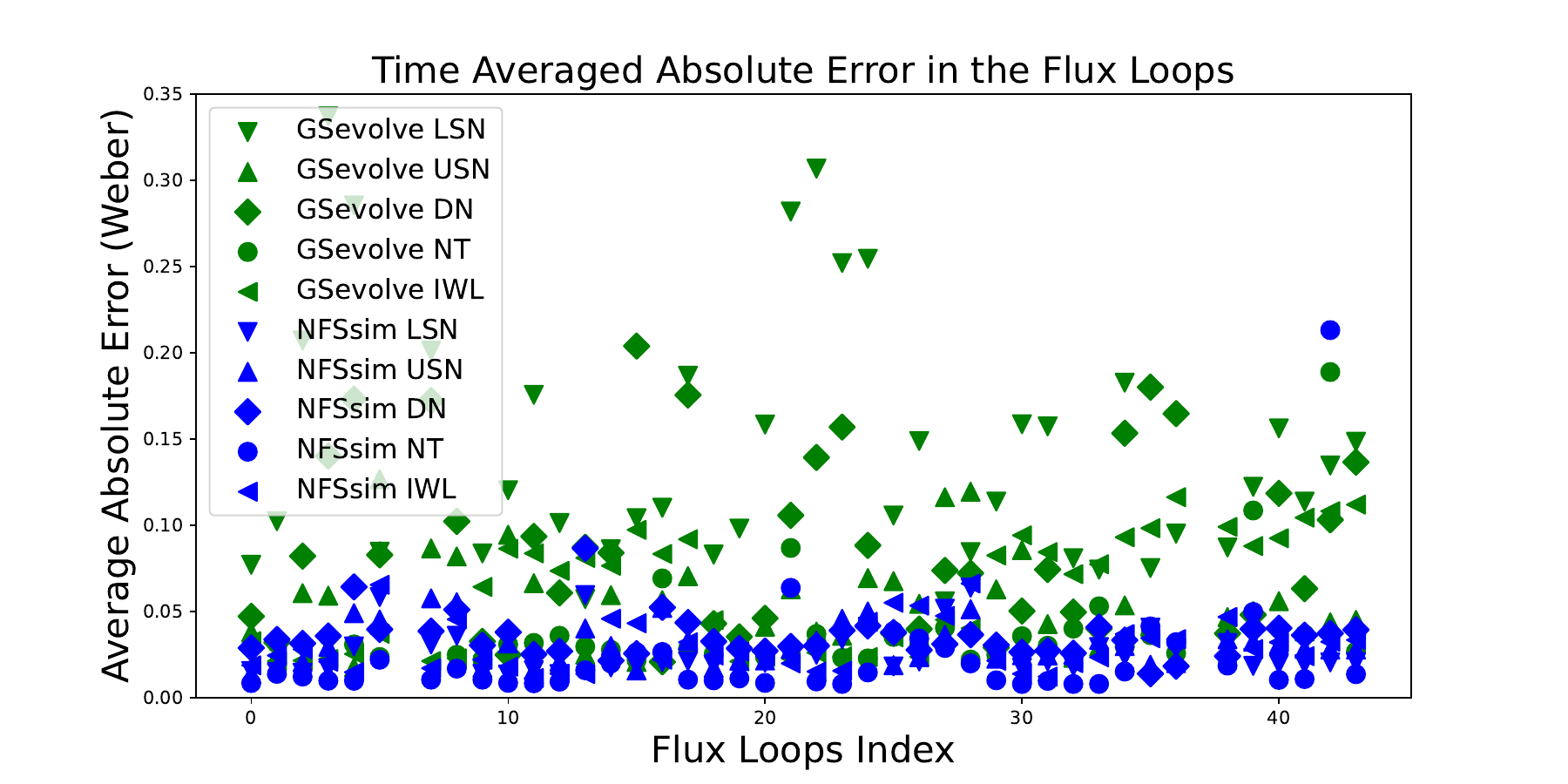}
\caption{This is a graph of the time-averaged absolute error in the flux loop sensors. Note that to make a fair comparison, the GSevolve flux loop data is normalized to remove offset from the time before NSFsim started running and the time average for the GSevolve data is only over the period of time when NSFsim is active. The missing columns are due to erroneous flux loops that were turned off during the shot as they were deemed unreliable. See the appendix for details on the Flux Loops Index.}
\label{FluxLoopsGraph}
\end{figure}

\begin{figure}[!ht]
\includegraphics[width=.48\textwidth]{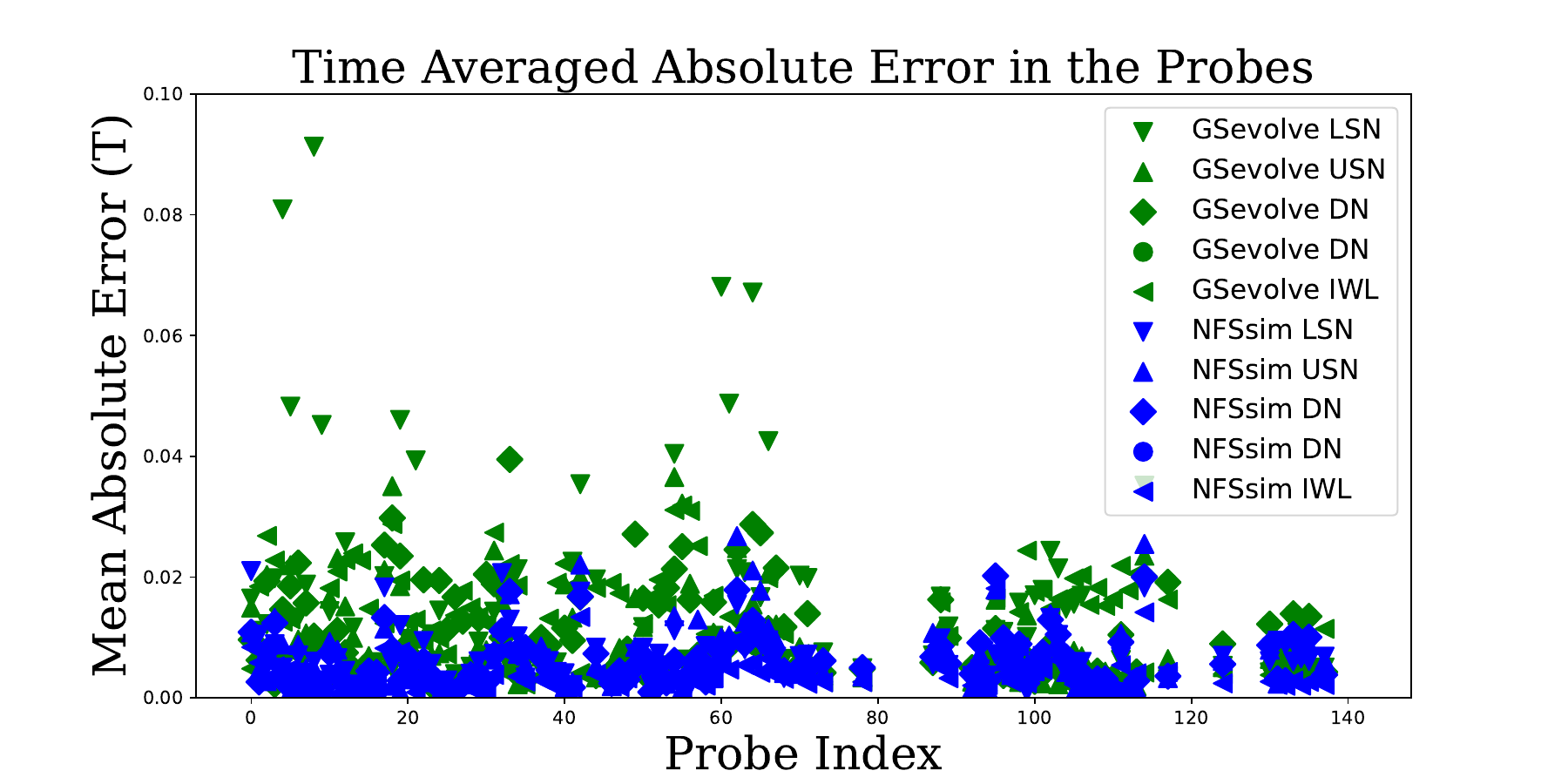}
\caption{This is a graph of the time-averaged absolute error in the magnetic probes. Unlike with the flux loop data, this data is not normalized as there was no significant offset. The GSevovle data is averaged over the time when NSFsim was active. The missing columns are due to erroneous probes that were turned off during the shot as they were deemed unreliable. See the appendix for details on the Probe Index.}
\label{MagProbesGraph}
\end{figure}

The magnetic diagnostic tools in the DIII-D tokamak are pivotal to understanding the status of the plasma, the shape itself is estimated from EFIT's calculations using flux loop and magnetic probe data. Shape control relies entirely on these diagnostic measurements and as such, their accuracy is paramount to controlling the plasma; throughout DIII-D's campaigns, flux loops and probes can be determined to be faulty and turned off to prevent erroneous data from disrupting the EFIT and thusly the plasma control (this can be seen in Figures \ref{FluxLoopsGraph} and \ref{MagProbesGraph} from their missing columns). \cite{ferron1998real}

Both NSFsim and GSevolve simulate these magnetic sensors for the purpose of emulating the best available data in a live DIII-D shot. Figures \ref{FluxLoopsGraph} and \ref{MagProbesGraph} are presented to give an idea of the qualitative performance of NSFsim relative to GSevolve for how accurate its probes and flux loops are over time; their simulated data is compared against the actual sensor data recorded in the shot they are recreating. The plots are presented this way to simplify down 142 probe and 44 flux loop data sets across 5 simulations done with both GSevolve and NSFsim in a way that can convey the accuracy of the NSFsim magnetic diagnostics relative to GSevolve's which are tried and true.


\subsection{NSFsim's Accuracy}
Up to this point the accuracy of NSFsim appears quite competitive with GSevolve's. The average absolute error in both the flux loops and magnetic probes show noticeably better consistency than in GSevolve, while the flux at the magnetic axis and plasma boundary doesn't show a clear winner. Both programs incur numerical error from the nature of doing numerical integration of differential equations, but they both have separate significant sources of error. GSevolve operates under active control conditions, simulating a live shot environment where the management of vertical instability can influence magnetic sensor data, potentially introducing fluctuations in contour alignment. NSFsim's source of non-numerical error is its inclusion of transport dynamics that haven't been fully validated with DIII-D; this is apparent in the graph of the flux at the magnetic axis and boundary where systemic error arises from the presence of transport effects that aren't accounted for while running NSFsim in reproduction mode, not trying to solve the transport problem.
\section{Conclusion}
NSFsim's first phase of development as a DIII-D plasma simulator is complete as it has successfully shown to recreate plasma shape, poloidal flux contours, and accurately simulate the magnetic diagnostic systems at DIII-D in comparison to real data and GSevovle data. The completion of the development opens up NSFsim as part of Fusion Twin Platform \cite{PlatformLink} that offers ease of use, fast plasma shape solution for scenario development, and offers transport mechanisms for modeling more complex plasma.

Future work on the NSFsim simulator involves deeper work on the modeling and validation of the available transport mechanisms at play in DIII-D. This will both minimize the errors that arise from high beta transport effects and expand NSFsim's utility for plasma study under a wider range of plasma scenarios. Additionally, this validation enables ongoing efforts to develop machine learning-based controllers in NSFsim for DIII-D.
\section{Acknowledgement}
This material is based upon work supported by the U.S. Department of Energy, Office of Science, Office of Fusion Energy Sciences, using the DIII-D National Fusion Facility, a DOE Office of Science user facility, under Award DE-FC02-04ER54698 and Next Step Fusion S.a.r.l. with UCSD staff supported by Next Step Fusion S.a.r.l.  The authors would like to thank Wilkie Choi, Himank Anand, and David Eldon for fruitful discussions.
\section{Disclaimer}
This report was prepared as an account of work sponsored by an agency of the United States Government. Neither the United States Government nor any agency thereof, nor any of their employees, makes any warranty, express or implied, or assumes any legal liability or responsibility for the accuracy, completeness, or usefulness of any information, apparatus, product, or process disclosed, or represents that its use would not infringe privately owned rights. Reference herein to any specific commercial product, process, or service by trade name, trademark, manufacturer, or otherwise does not necessarily constitute or imply its endorsement, recommendation, or favoring by the United States Government or any agency thereof. The views and opinions of authors expressed herein do not necessarily state or reflect those of the United States Government or any agency thereof.
\addcontentsline{toc}{chapter}{Bibliography}
\bibliographystyle{plain} 


\begin{thebibliography}{10}

\bibitem{buneman1969compact}
C~Buneman.
\newblock A compact non-iterative poisson solver.
\newblock {\em Stanford Univ. Inst. Plasma Res. Rep.}, 294, 1969.

\bibitem{burrell2002quiescent}
KH~Burrell, Max~E Austin, DP~Brennan, JC~DeBoo, EJ~Doyle, P~Gohil, CM~Greenfield, RJ~Groebner, LL~Lao, TC~Luce, et~al.
\newblock Quiescent h-mode plasmas in the diii-d tokamak.
\newblock {\em Plasma Physics and Controlled Fusion}, 44(5A):A253, 2002.

\bibitem{degrave2022magnetic}
Jonas Degrave, Federico Felici, Jonas Buchli, Michael Neunert, Brendan Tracey, Francesco Carpanese, Timo Ewalds, Roland Hafner, Abbas Abdolmaleki, Diego de~Las~Casas, et~al.
\newblock Magnetic control of tokamak plasmas through deep reinforcement learning.
\newblock {\em Nature}, 602(7897):414--419, 2022.

\bibitem{degtyarev1969flow}
Lev~Markovich Degtyarev and AP~Favorskii.
\newblock Flow variant of the sweep method for difference problems with strongly varying coefficients.
\newblock {\em USSR Computational Mathematics and Mathematical Physics}, 9(1):285--294, 1969.

\bibitem{degtyarev1985inverse}
LM~Degtyarev and VV~Drozdov.
\newblock An inverse variable technique in the mhd-equilibrium problem.
\newblock {\em Computer Physics Reports}, 2(7):341--387, 1985.

\bibitem{eldon2020high}
D~Eldon, AW~Hyatt, B~Covele, N~Eidietis, HY~Guo, DA~Humphreys, AL~Moser, B~Sammuli, and ML~Walker.
\newblock High precision strike point control to support experiments in the diii-d small angle slot divertor.
\newblock {\em Fusion Engineering and Design}, 160:111797, 2020.

\bibitem{favez2001comparing}
JY~Favez, RR~Khayrutdinov, JB~Lister, and VE~Lukash.
\newblock Comparing tcv experimental vde responses with dina code simulations.
\newblock {\em Plasma physics and controlled fusion}, 44(2):171, 2001.

\bibitem{ferron1998real}
JR~Ferron, ML~Walker, LL~Lao, HE~St John, DA~Humphreys, and JA~Leuer.
\newblock Real time equilibrium reconstruction for tokamak discharge control.
\newblock {\em Nuclear fusion}, 38(7):1055, 1998.

\bibitem{khayrutdinov2001comparing}
RR~Khayrutdinov, JB~Lister, VE~Lukash, and JP~Wainwright.
\newblock Comparing dina code simulations with tcv experimental plasma equilibrium responses.
\newblock {\em Plasma Physics and Controlled Fusion}, 43(3):321, 2001.

\bibitem{khayrutdinov1993studies}
RR~Khayrutdinov and VE~Lukash.
\newblock Studies of plasma equilibrium and transport in a tokamak fusion device with the inverse-variable technique.
\newblock {\em Journal of Computational Physics}, 109(2):193--201, 1993.

\bibitem{kim2009full}
SH~Kim, JF~Artaud, V~Basiuk, V~Dokuka, RR~Khayrutdinov, JB~Lister, and VE~Lukash.
\newblock Full tokamak discharge simulation of iter by combining dina-ch and cronos.
\newblock {\em Plasma Physics and Controlled Fusion}, 51(10):105007, 2009.

\bibitem{lao1985reconstruction}
LL~Lao, H~St John, RD~Stambaugh, AG~Kellman, and W~Pfeiffer.
\newblock Reconstruction of current profile parameters and plasma shapes in tokamaks.
\newblock {\em Nuclear fusion}, 25(11):1611, 1985.

\bibitem{lazarus1994role}
EA~Lazarus, AW~Hyatt, and TH~Osborne.
\newblock The role of shaping in achieving high performance in diii-d.
\newblock Technical report, General Atomics, 1994.

\bibitem{liang2018rllib}
Eric Liang, Richard Liaw, Robert Nishihara, Philipp Moritz, Roy Fox, Ken Goldberg, Joseph~E. Gonzalez, Michael~I. Jordan, and Ion Stoica.
\newblock {RLlib}: Abstractions for distributed reinforcement learning.
\newblock In {\em International Conference on Machine Learning ({ICML})}, 2018.

\bibitem{lukash2007influence}
VE~Lukash, AB~Mineev, and D~Kh Morozov.
\newblock Influence of plasma opacity on current decay after disruptions in tokamaks.
\newblock {\em Nuclear fusion}, 47(11):1476, 2007.

\bibitem{miyamoto2014inter}
S~Miyamoto, A~Isayama, I~Bandyopadhyay, SC~Jardin, RR~Khayrutdinov, VE~Lukash, Y~Kusama, and M~Sugihara.
\newblock Inter-code comparison benchmark between dina and tsc for iter disruption modelling.
\newblock {\em Nuclear Fusion}, 54(8):083002, 2014.

\bibitem{nelson2023robust}
Andrew~O Nelson, Lothar Schmitz, Carlos Paz-Soldan, Kathreen~E Thome, TB~Cote, N~Leuthold, F~Scotti, Max~E Austin, Alan Hyatt, and Thomas Osborne.
\newblock Robust avoidance of edge-localized modes alongside gradient formation in the negative triangularity tokamak edge.
\newblock {\em Physical Review Letters}, 131(19):195101, 2023.

\bibitem{Raffin2021SB3}
Antonin Raffin, Ashley Hill, Adam Gleave, Anssi Kanervisto, Maximilian Ernestus, and Noah Dormann.
\newblock Stable-baselines3: Reliable reinforcement learning implementations.
\newblock {\em Journal of Machine Learning Research}, 22(268):1--8, 2021.

\bibitem{rutherford2024manta}
Grant Rutherford, Haley~S Wilson, Audrey Saltzman, David Arnold, John~Leland Ball, Stuart Royce~Sands Benjamin, Rachel Bielajew, Nikolai de~Boucaud, Miguel Calvo~Carrera, Rian Chandra, et~al.
\newblock Manta: A negative-triangularity nasem-compliant fusion pilot plant.
\newblock {\em Plasma Physics and Controlled Fusion}, 2024.

\bibitem{Samarskiy1975}
Iu.~P. Samarskii, A. A. ;~Popov.
\newblock “difference methods for solving problems of gas dynamics /2nd revised and enlarged edition”.
\newblock {\em Moscow, Izdatel'stvo Nauka}, 1980.

\bibitem{PlatformLink}
Next Step~Fusion S.a.r.l.
\newblock Fusion twin platform.

\bibitem{tamai2002runaway}
H~Tamai, R~Yoshino, S~Tokuda, G~Kurita, Y~Neyatani, M~Bakhtiari, RR~Khayrutdinov, VE~Lukash, MN~Rosenbluth, et~al.
\newblock Runaway current termination in jt-60u.
\newblock {\em Nuclear fusion}, 42(3):290, 2002.

\bibitem{towers2024gymnasium}
Mark Towers, Ariel Kwiatkowski, Jordan Terry, John~U. Balis, Gianluca~De Cola, Tristan Deleu, Manuel Goulão, Andreas Kallinteris, Markus Krimmel, Arjun KG, Rodrigo Perez-Vicente, Andrea Pierré, Sander Schulhoff, Jun~Jet Tai, Hannah Tan, and Omar~G. Younis.
\newblock Gymnasium: A standard interface for reinforcement learning environments, 2024.

\bibitem{welander2019closed}
Anders Welander, Erik Olofsson, Brian Sammuli, Michael~L Walker, and Bingjia Xiao.
\newblock Closed-loop simulation with grad-shafranov equilibrium evolution for plasma control system development.
\newblock {\em Fusion Engineering and Design}, 146:2361--2365, 2019.

\bibitem{xue2019hot}
L~Xue, GY~Zheng, XR~Duan, YQ~Liu, GT~Hoang, JX~Li, VN~Dokuka, VE~Lukash, and RR~Khayrutdinov.
\newblock Hot vde investigation of the negative triangularity plasmas based on hl-2m tokamak.
\newblock {\em Fusion Engineering and Design}, 143:48--58, 2019.

\bibitem{xue2015simulation}
Lei Xue, Xu-Ru Duan, Guo-Yao Zheng, Yue-Qiang Liu, Shi-Lei Yan, VV~Dokuka, RR~Khayrutdinov, and VE~Lukash.
\newblock Simulation of plasma disruptions for hl-2m with the dina code.
\newblock {\em Chinese Physics Letters}, 32(6):065203, 2015.

\end{thebibliography}

\begin{appendices}
\onecolumn
\section{Magnetic Sensor Index}

\begin{table}[!ht]
\centering
\begin{tabular}{ |p{2.5cm}||p{2.5cm}|p{2.5cm}|p{2.5cm}|p{2.5cm}|  }
\hline
\multicolumn{5}{|c|}{(Index Number) Flux Loop Sensor Name} \\
\hline
(1) PSF1A & (2) PSF2A & (3) PSF3A & (4) PSF4A & (5) PSF5A \\
\hline
(6) PSF6NA & (7) PSF7NA & (8) PSF8A & (9) PSF9A & (10) PSF1B \\
\hline
(11) PSF2B & (12) PSF3B & (13) PSF4B & (14) PSF5B & (15) PSF6NB \\
\hline
(16) PSF7NB & (17) PSF8B & (18) PSF9B & (19) PSI11M & (20) PSI12A \\
\hline
(21) PSI23A & (22) PSI34A & (23) PSI45A & (24) PSI58A & (25) PSI9A \\
\hline
(26) PSF7FA & (27) PSI7A & (28) PSF6FA & (29) PSI6A & (30) PSI12B \\
\hline
(31) PSI23B & (32) PSI34B & (33) PSI45B & (34) PSI58B & (35) PSI9B \\
\hline
(36) PSF7FB & (37) PSI7B & (38) PSF6FB & (39) PSI6B & (40) PSI89FB \\
\hline
(41) PSI89NB & (42) PSI1L & (43) PSI2L & (44) PSI3L &  \\
\hline
\end{tabular}
\caption{Table of Flux Loop sensors and their name with the index number corresponding to the index in Figure \ref{FluxLoopsGraph}}
\label{FL_Table_Index}
\end{table}

\begin{table}[!ht]
\centering
\begin{tabular}{ |p{2.8cm}||p{2.8cm}|p{2.8cm}|p{2.8cm}|p{2.8cm}|  }
\hline
\multicolumn{5}{|c|}{(Index Number) Magnetic Probe Sensor Name} \\
\hline
(1) MPI11M322 & (2) MPI1A322 & (3) MPI2A322 & (4) MPI3A322 & (5) MPI4A322 \\
\hline
(6) MPI5A322 & (7) MPI8A322 & (8) MPI89A322 & (9) MPI9A322 & (10) MPI79FA322 \\
\hline
(11) MPI79NA322 & (12) MPI7FA322 & (13) MPI7NA322 & (14) MPI67A322 & (15) MPI6FA322 \\
\hline
(16) MPI6NA322 & (17) MPI66M322 & (18) MPI1B322 & (19) MPI2B322 & (20) MPI3B322 \\
\hline
(21) MPI4B322 & (22) MPI5B322 & (23) MPI8B322 & (24) MPI89B322 & (25) MPI9B322 \\
\hline
(26) MPI79B322 & (27) MPI7FB322 & (28) MPI7NB322 & (29) MPI67B322 & (30) MPI6FB322 \\
\hline
(31) MPI6NB322 & (32) MPI2A067 & (33) MPI11M067 & (34) MPI2B067 & (35) MPI67A067 \\
\hline
(36) MPI66M067 & (37) MPI67B067 & (38) MPI1A139 & (39) MPI2A139 & (40) MPI3A139 \\
\hline
(41) MPI4A139 & (42) MPI5A139 & (43) MPI79A147 & (44) MPI7NA142 & (45) MPI67A157 \\
\hline
(46) MPI6FA142 & (47) MPI6NA132 & (48) MPI6NA157 & (49) MPI66M157 & (50) MPI6NB157 \\
\hline
(51) MPI6FB142 & (52) MPI67B157 & (53) MPI7NB142 & (54) MPI79B142 & (55) MPI5B139 \\
\hline
(56) MPI4B139 & (57) MPI3B139 & (58) MPI2B139 & (59) MPI1B139 & (60) MPI1B157 \\
\hline
(61) MPI1U157 & (62) MPI2U157 & (63) MPI3U157 & (64) MPI4U157 & (65) MPI5U157 \\
\hline
(66) MPI6U157 & (67) MPI7U157 & (68) DSL1U180 & (69) DSL2U180 & (70) DSL3U180 \\
\hline
(71) DSL4U157 & (72) DSL5U157 & (73) DSL6U157 & (74) MPI66M127 & (75) MPI66M132 \\
\hline
(76) MPI66M137 & (77) MPI66B137 & (78) MPI6NB137 & (79) MPI66M307 & (80) MPI66M312 \\
\hline
(81) MPI6NA312 & (82) MPI66B312 & (83) MPI6NB312 & (84) MPI1L020 & (85) MPI2L020 \\
\hline
(86) MPI1L050 & (87) MPI1L110 & (88) MPI1L180 & (89) MPI2L180 & (90) MPI3L180 \\
\hline
(91) MPI1L230 & (92) MPI1L320 & (93) MPI66M020 & (94) MPI66M097 & (95) MPI66M200 \\
\hline
(96) MPI66M247 & (97) MPI66M277 & (98) MPI66M340 & (99) MPI67A022 & (100) MPI67A037 \\
\hline
(101) MPI67A052 & (102) MPI67A082 & (103) MPI67A097 & (104) MPI67A142 & (105) MPI67A217 \\
\hline
(106) MPI67A277 & (107) MPI67A337 & (108) MPI67B022 & (109) MPI67B037 & (110) MPI67B052 \\
\hline
(111) MPI67B097 & (112) MPI67B217 & (113) MPI67B277 & (114) MPI67B337 & (115) MPI79A072 \\
\hline
(116) MPI79A222 & (117) MPI79A272 & (118) MPI79B067 & (119) MPI79B217 & (120) MPI79B277 \\
\hline
(121) MPI5A199 & (122) MPI4A199 & (123) MPI3A199 & (124) MPI2A199 & (125) MPI1A199 \\
\hline
(126) MPI1B199 & (127) MPI2B199 & (128) MPI3B199 & (129) MPI4B199 & (130) MPI5B199 \\
\hline
(131) MPI1A011 & (132) MPI1A049 & (133) MPI1A109 & (134) MPI1A244 & (135) MPI1A274 \\
\hline
(136) MPI1A341 & (137) MPI1B011 & (138) MPI1B049 & (139) MPI1B109 & (140) MPI1B244 \\
\hline
(141) MPI1B274 & (142) MPI1B341 &  &  &  \\
\hline
\end{tabular}
\caption{Table of Magnetic Probes and their name with the index number corresponding to the index in Figure \ref{MagProbesGraph}}
\label{MagProbe_Table_Index}
\end{table}

\end{appendices}

\end{document}